\documentclass[twocolumn,aps,prl,draft]{revtex4}
\usepackage{graphicx}
\usepackage{bm}
\begin{document}
\title{Onset and universality of drag reduction in the turbulent 
Kolmogorov flow}
\author{Guido Boffetta$^1$, Antonio Celani$^2$, and Andrea Mazzino$^{3,4}$}
\affiliation{$^1$ Dipartimento di Fisica Generale and INFM, 
Universit\`a degli Studi di Torino, Via Pietro Giuria 1, 
10125, Torino, Italy \\
$^2$ CNRS, INLN, 1361 Route des Lucioles, 06560 Valbonne, 
France \\
$^3$ ISAC/CNR, Lecce Section, I--73100 Lecce, Italy\\
$^4$ INFM-Physics Department, Genova University, Via Dodecaneso 33, 
16146 Genova, Italy}
\date{\today}
\begin{abstract}
We investigate the phenomenon of drag reduction in a viscoelastic fluid
model of dilute polymer solutions. By means of direct numerical
simulations of the three-dimensional turbulent Kolmogorov flow we show that
drag reduction takes place above a critical Reynolds number $Re_{c}$.
An explicit expression for the dependence of $Re_{c}$ on
polymer elasticity and diffusivity is derived. The values of 
the drag coefficient obtained for different fluid parameters collapse
onto a universal curve when plotted as a function of the
rescaled Reynolds number $Re/Re_{c}$.
The analysis of the momentum budget allows to gain some insight
on the physics of drag reduction, and suggests the existence of 
a maximum drag reduction asymptote for this flow.
\end{abstract}
\maketitle 
When a viscous fluid is kept in motion by some external driving, 
a mean flow is established: the ratio between the work made
by the force and the kinetic energy carried by the mean flow
is called the {\em drag coefficient}, or friction factor.
This dimensionless number measures the power that has to be supplied to
the fluid to maintain a given throughput. When the flow is laminar,
the drag coefficient is inversely proportional to the Reynolds number. 
Upon increasing the intensity of the
applied force the flow eventually becomes turbulent, and the drag
coefficient becomes approximately independent of the Reynolds number
\cite{R1883}, therefore substantially larger than in the viscous case.    

In 1949 the British chemist Toms reported that the turbulent drag 
could be reduced by up to 80\% through the addition of minute amounts
(few tenths of p.p.m. in weight) of long-chain soluble polymers
to water. This observation triggered an enormous experimental activity 
to characterize this phenomenon (see, e.g.,
\cite{L69,V75,MC92,NH95,SW00}).
In spite of these efforts, no fully satisfactory theory of drag
reduction is available yet. However, a recent breakthrough has been
the observation of drag reduction in numerical simulations of the
turbulent channel flow of viscoelastic fluids \cite{SBH97}.
Most of the features of experimental flows of dilute
polymer solutions are successfully reproduced by these models, 
even at the quantitative level \cite{PBNHVH03}.
Despite these advances, the understanding of drag reduction in the
experimentally relevant geometry of pipe or channel flow is still hindered
by the complexity of these flows already at the Newtonian
level, i.e. in the absence of polymers \cite{DCLPP03}.
This consideration motivated us to investigate 
simpler geometries in the hope that this may shed some
light on the basic physical mechanisms of drag reduction
(see, e.g., Ref.~\cite{DCBP02}).

In this Letter we present the results of an extensive numerical
investigation of the viscoelastic turbulent Kolmogorov flow.
This system has several analogies with the turbulent channel
flow, while its main distinctive trait is the absence of material boundaries.
Notwithstanding this major difference we will show that 
drag reduction takes place in the Kolmogorov flow as well. 
Furthermore, we observe striking quantitative similarities 
with experimental results in wall-bounded flows: this
points to the conclusion that the basic physical 
mechanisms of drag reduction be substantially independent of 
the detailed structure of the flow.

To describe the dynamics of a dilute polymer solution we 
adopt the linear viscoelastic model (Oldroyd-B) \cite{BCAH87}  
\begin{equation}
\partial_t {\bm u} + ({\bm u}\cdot{\bm \nabla}) {\bm u}
=-{\bm \nabla p} + \nu_0 {\Delta} {\bm u} + \frac{2 \eta\,\nu_0}{\tau} 
{\bm \nabla}\cdot{\bm \sigma}
+ {\bm F}, 
\label{eq:1}
\end{equation}
\begin{equation}
\partial_t {\bm \sigma} + ({\bm u}\cdot{\bm \nabla}) {\bm \sigma}
\!=\!({\bm \nabla \bm u})^T\! \cdot {\bm \sigma} + {\bm \sigma}\! 
\cdot ({\bm \nabla \bm u})
-2\frac{{\bm \sigma}-{\bm 1}}{\tau} + \kappa \Delta {\bm \sigma}.
\label{eq:2}
\end{equation}
The velocity field ${\bm u}$ is incompressible,
the symmetric matrix  ${\bm \sigma}$ is
the conformation tensor of polymer 
molecules, and its trace $\textrm{tr}{\bm \sigma}$ is a measure of 
their elongation. 
The parameter $\tau$ is the (slowest) polymer relaxation time.
The matrix of velocity gradients is defined as 
$({\bm \nabla \bm u})_{ij}=\partial_i u_j$ and ${\bm 1}$ is the unit tensor.
The solvent viscosity is denoted by $\nu_0$ and
$\eta$ is the zero-shear contribution of polymers to the total solution
viscosity $\nu=\nu_0(1+\eta)$. The parameter $\eta$ is proportional
to the polymer concentration. The diffusive term $\kappa \Delta {\bm
  \sigma}$ is added to prevent numerical instabilities \cite{SB95}.
The constant forcing ${\bm F}=(F\cos(z/L),0,0)$ 
maintains the system in a statistically stationary state 
characterized by a mean flow $\langle {\bm u} \rangle$.
Due to the symmetries of ${\bm F}$, the only nonzero component 
of the mean velocity is 
$\langle u_x \rangle$: it depends
on the shear coordinate $z$ alone, 
vanishes at $z=\pm (\pi/2) L$, and is even under reflections
$z \to -z$. Its value at $z=0$,
$\langle u_x \rangle_{z=0}$, will be denoted by $U$. Finally, we
establish a short glossary between the Kolmogorov flow and the channel flow:
$F$ plays the role of the pressure gradient, $\pi L$ is analogous to 
the channel height, and $U$ is equivalent to the centerline velocity.

In this framework, we have performed a series of 
numerical integrations of eqs.~(\ref{eq:1}) and (\ref{eq:2}) 
for a set of values of forcing
intensity $F$, at fixed $\nu$, both for the Newtonian
and the viscoelastic case. 
Comparing results at a given $F$ is 
equivalent to keeping an imposed pressure gradient -- therefore a fixed
wall-shear stress -- in channel flow experiments
(see, e.g. Ref.~\cite{LT88}).
We have measured the 
mean profiles of several relevant observables, including the average velocity
$\langle u_x \rangle$, the turbulent shear stress (Reynolds stress)
$\langle u_x u_z \rangle$, and the mean polymer stress 
$2\nu_0\eta\tau^{-1}\langle \sigma_{xz}\rangle$. 
The mean flow is accurately described by the sinusoidal
profile $\langle  u_x \rangle = U \cos(z/L)$, both
in the Newtonian and in the viscoelastic flow \cite{profiles}.
\begin{figure}[t]
\hspace*{-0.5cm}\includegraphics[draft=false,scale=0.65]{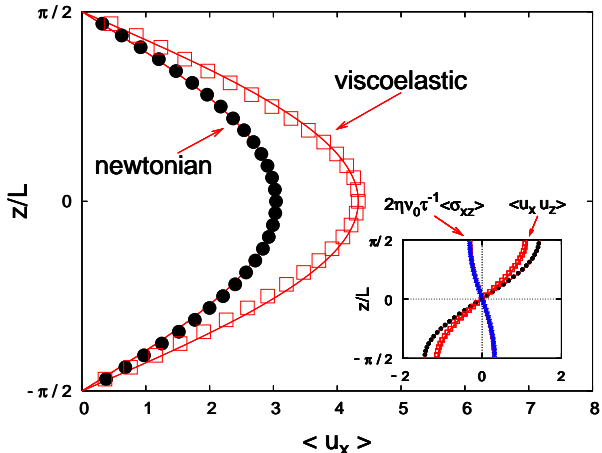}
\caption{Mean velocity profiles for 
a Newtonian ($\eta=0$) and a viscoelastic simulation ($\eta=0.3$, $El=0.019$) 
at given forcing amplitude $F=1.5$. The measured profiles are undistinguishable
from $\langle u_x \rangle = U \cos(z/L)$ (full lines) in both cases.
The effect of elasticity
is to increase the peak value $U$ with respect to the Newtonian case:
in the present case this corresponds to a reduction of the drag coefficient, 
defined in eq.~(\ref{eq:3}), of about $40\%$.
In the inset, the profiles of the
Reynolds stress $\langle u_x u_z \rangle= S \sin(z/L)$ 
and the mean polymer
stress $2\nu_0\eta\tau^{-1}\langle \sigma_{xz}\rangle=-T\sin(z/L)$. 
In this case 
the Reynolds stress is reduced upon polymer addition to approximately 
$70\%$ of its Newtonian value, consistently with
experimental results at comparable drag reduction \cite{WWL87}. 
The "missing" turbulent 
shear stress is compensated by the contribution of the polymer stress:
the sum of $S$ and $T$ is  equal to $F$ in both the 
Newtonian and viscoelastic case.
Data result from the numerical integration
of eqs. (\protect\ref{eq:1}) and (\protect\ref{eq:2}) in
a periodic cube of side $2\pi$ by means of a 
fully dealiased pseudospectral code with $64^3$ 
collocation points. 
The mean flow lengthscale is $L=1$ and the viscosity is $\nu=0.015625$.
Starting from an initial configuration with a 
small amount of energy on the smallest modes, after the system
evolved into a statistically stationary state, 
time averages over 100 to 1000 eddy-turnover times
have been performed to obtain the mean velocity profiles.}
\label{fig:0}
\end{figure}
However, as shown in Fig.~\ref{fig:0}, 
in the latter case the centerline velocity $U$ is definitely larger:
this is the hallmark of drag reduction.
It has to be remarked that -- at variance with wall-bounded flows 
where drag reduction is always 
accompanied by a structural change in the profile (see e.g. Ref.~\cite{V75}) --
in the Kolmogorov flow the increase in throughput
takes place just by means of an overall rescaling of the mean velocity.
This is due to the different boundary conditions: in channel flows,
the profile in the viscous sublayer is left 
unchanged upon polymer addition
while the bulk flow increases substantially. 
This requires a reshaping of the mean profile, 
that takes actually place through the increase of the extent
of the buffer region (see e.g. Ref.~\protect\cite{LT88}).
In the Kolmogorov flow there is no constraint on velocity profiles,
and drag reduction does not necessarily entail their structural change.

To quantify the effect of viscoelasticity on the mean flow, 
we have defined the drag coefficient as
\begin{equation}
f=\frac{FL}{U^2}\;,
\label{eq:3}
\end{equation}
and measured its dependence on the Reynolds number 
$Re=UL/\nu$ \cite{Rey-num}.
For $Re<\sqrt{2}$ the flow is 
laminar with mean velocity $U=FL^2/\nu$, giving a drag coefficient 
$f=Re^{-1}$. At $Re \gtrsim 50$ the system is already 
in a fully developed turbulent
state. 
\begin{figure}[b]
\centering
\includegraphics[draft=false, scale=0.73]{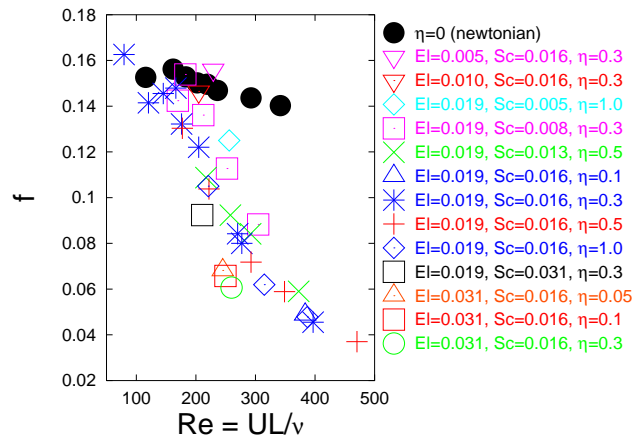}
\caption{The drag coefficient for different viscoelastic fluid
  parameters. Data have been collected from numerical simulations
  at different forcing amplitudes $F$ and viscoelastic parameters $\eta$,
  $\tau$, $\kappa$.
The statistical error in the determination of $f$ and $Re$  
is of the order of the symbols' size.}
\label{fig:1} 
\end{figure}
For a Newtonian fluid,
numerical data show that the drag coefficient is 
approximately independent of $Re$ (see Fig.~\ref{fig:1}). 
This behaviour agrees with the following classical Kolmogorov argument:
since the average energy input
$\epsilon = FU/2$
scales as $\epsilon = \frac{\beta}{2}\, U^{3}/L$
in fully developed turbulence, 
eq.~(\ref{eq:3}) yields a constant drag coefficient $f=\beta$.
The Newtonian momentum budget gives
$F_x=\partial_z \langle u_x u_z\rangle$ 
(the viscous contribution being negligible) and therefore a Reynolds stress 
$\langle u_x u_z \rangle=S \sin(z/L)$ with $S=\beta U^2$. 
For the turbulent Kolmogorov flow, $\beta\simeq 0.15$.

\begin{figure}[b]
\centering
\includegraphics[draft=false, scale=0.17]{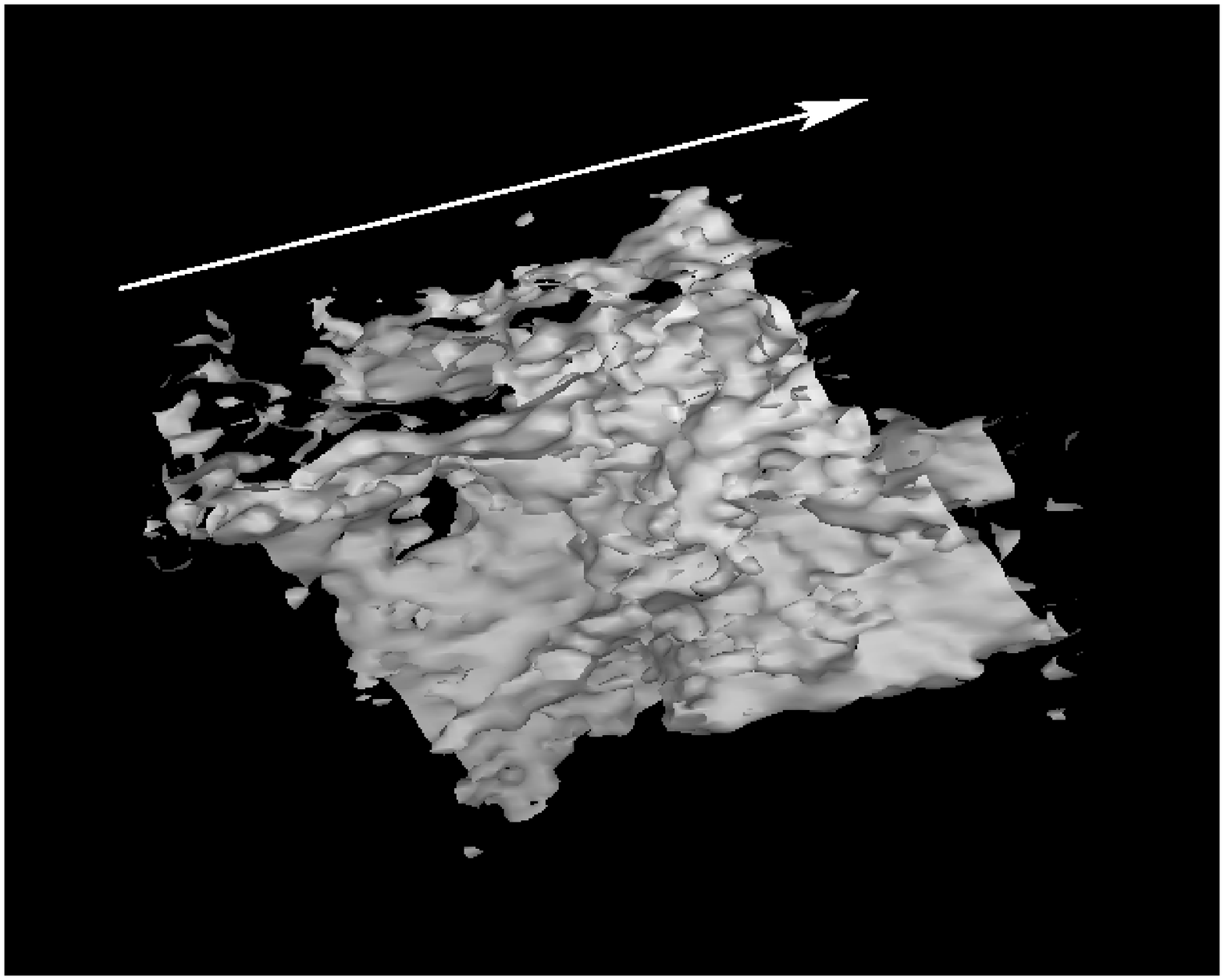}
\includegraphics[draft=false, scale=0.17]{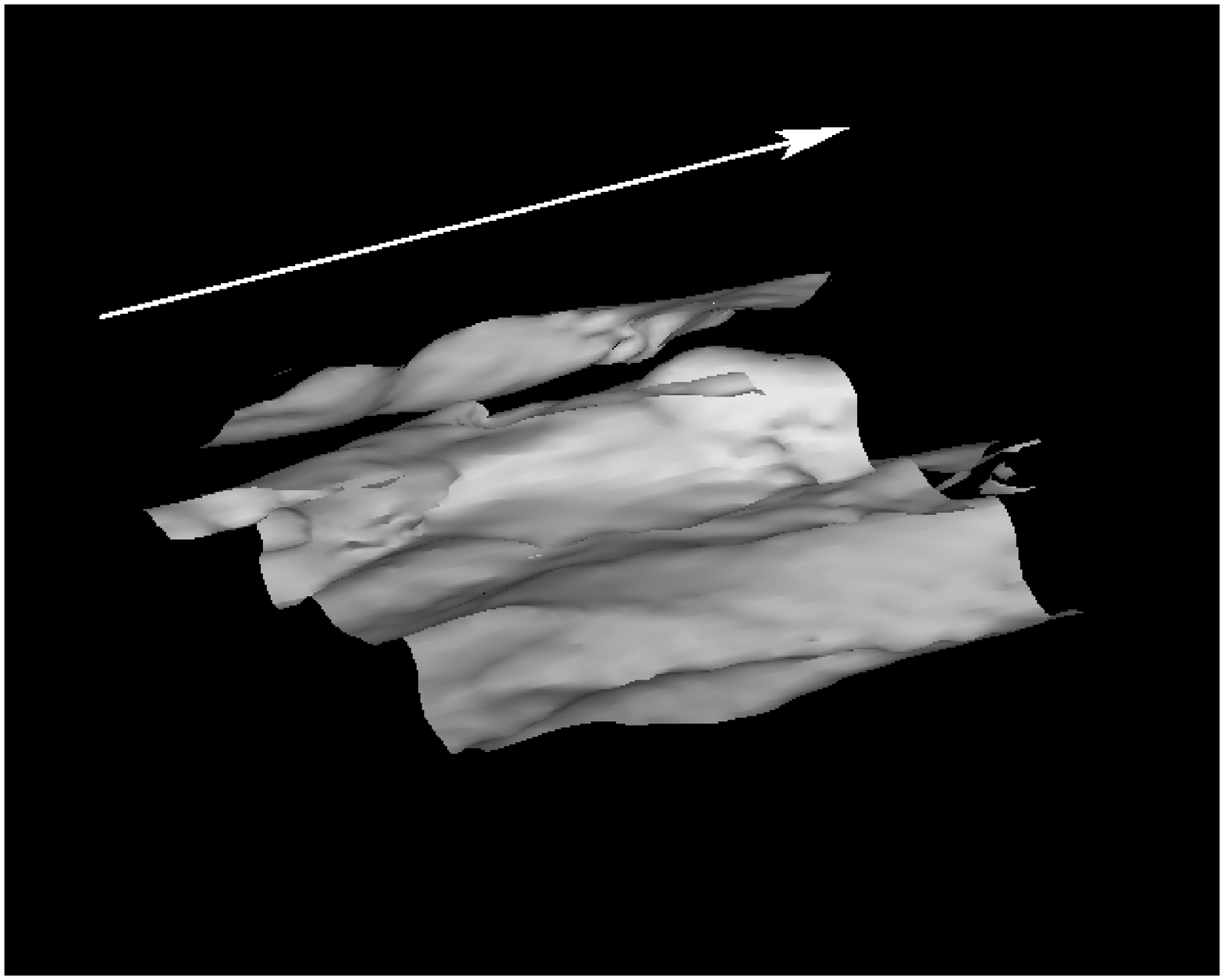}
\caption{Snapshots of the isosurfaces $u_x(x,y,z)=U$
for a Newtonian (left) and a viscoelastic simulation
(right, $El=0.019$, $Sc=0.016$, $\eta=0.5$). The Reynolds number is 
$Re \approx 350$. The arrows show the direction of the mean flow.
Small-scale turbulent
fluctuations, responsible for kinetic energy dissipation, 
are suppressed in the viscoelastic case.
A high-speed streak in the form a 
streamwise oriented tube is visible in the viscoelastic case (right).}
\label{fig:2} 
\end{figure}

When polymers are added $f$ may be reduced with respect to its
Newtonian value, depending on the polymer elasticity $El=\nu\tau/L^2$,
the Schmidt number $Sc=\nu/\kappa$, and the concentration $\eta$,
 as shown in Fig.~\ref{fig:1}. 
For the highest Reynolds number we can attain in our simulations
the friction factor is reduced by 75\%.
Drag reduction is accompanied by changes in the 
velocity field similar to those occurring in channel flow experiments
and simulations: the level of transverse fluctuations $\langle u_z^2 \rangle$
is reduced while
longitudinal fluctuations $\langle (u_x-\langle u_x \rangle)^2 \rangle $
increase
and high streamwise velocity streaks are observed
(see Fig.~\ref{fig:2}).  
Incidentally, we notice that drag reduction is observed at Reynolds numbers 
definitely smaller than the typical experimental values: 
this is possible thanks to the relatively high value of elasticity
utilized in our simulations. Comparable parameters have been used
in numerical simulations of the channel flow as well (see, e.g. 
Ref.~\cite{SBH97}), and produced a similar effect on the threshold 
for drag reduction. 
 
 From the inspection of Fig.~\ref{fig:1} we notice some systematic
trend:
at moderate Reynolds numbers ($Re \lesssim 200$) 
viscoelastic effects do not alter substantially 
the value of the
drag coefficient; at larger $Re$  
polymers with a higher elasticity 
are more effective as drag-reducing agents; conversely, polymers 
with higher diffusivity are less effective. 
To understand the variation of the drag coefficient with fluid parameters,
we sought a dependence of the form 
$f=\varphi(Re/Re_{c})$ where $Re_{c}(El,Sc,\eta)$ is 
the critical Reynolds number for the onset of drag reduction.
To obtain an explicit expression for $Re_{c}$ we 
need to extend the argument given in Ref.~\cite{BFL01} to the case 
of finite polymer diffusivity. The reasoning goes as follows:
for polymers to be substantially elongated, stretching must prevail
over elastic relaxation and diffusivity \cite{timecr}; 
at the onset, the terms appearing
in eq.~(\ref{eq:2}) must then satisfy
$(\nabla u)_c \sim 2/\tau + \kappa/L^2$; since the transition is incipient
we can estimate the typical 
velocity gradient as $(\epsilon_c/\nu)^{1/2}$,
and utilizing the expression $\epsilon_c \propto U_c^3/L$ we finally obtain
\begin{equation}
Re_{c} \propto \left( \frac{2}{El} + \frac{1}{Sc} \right)^{2/3} \;.
\label{eq:4}
\end{equation} 
For vanishing diffusivity we recover the result of Ref.~\cite{BFL01}.
Extracting the explicit dependence on polymer concentration,
we have $Re_c \propto (1+\eta)^{-2/3}$, 
which is compatible with the weak dependence on concentration 
found in experiments \cite{V75}.

\begin{figure}[b]
\centering
\includegraphics[draft=false, scale=0.73]{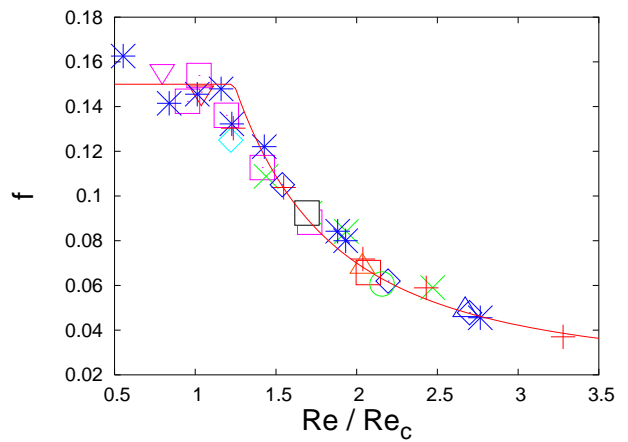}
\caption{The drag coefficient plotted as a function of the rescaled 
Reynolds number $Re/Re_{c}$. Symbols as in fig.~\ref{fig:1}. 
The full line is eq.~(\ref{eq:5}) with $\beta=0.15$, 
$\gamma=0.2$ and $\delta=0.02$.}
\label{fig:3} 
\end{figure}

In Fig.~\ref{fig:3} we present the same data as in Fig.~\ref{fig:1},
now plotted against the rescaled Reynolds number $Re/Re_{c}$.
The good quality of the collapse supports the validity of 
the relation $f=\varphi(Re/Re_{c})$. The function $\varphi$ is
universal with respect to the choice of fluid parameters. Its 
shape will be derived in the following, with the aid of simple 
assumptions, starting from the equation for momentum conservation 
(see Ref.~\cite{LPPT03} for a similar approach to wall-bounded flows). 
\begin{figure}[t]
\centering
\includegraphics[draft=false, scale=0.65]{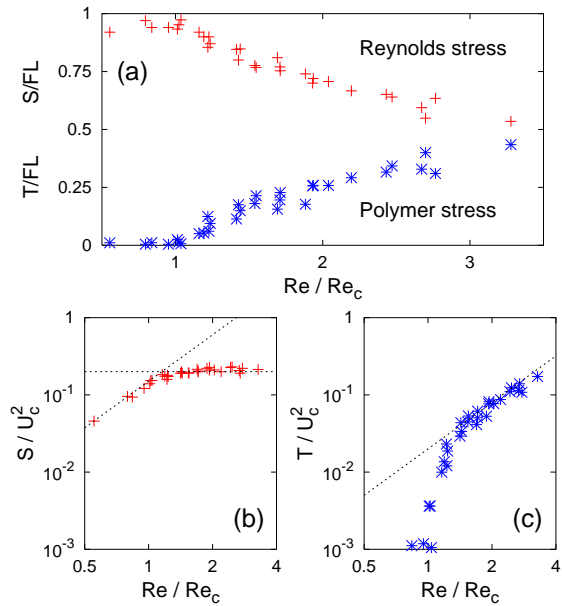}
\caption{$(a)$: Peak values of the Reynolds stress 
$\langle u_x u_z \rangle = S \sin(z/L)$,
and of the polymer stress 
$2 \eta \nu_0 \tau^{-1} \langle \sigma_{xz} \rangle = - T \sin(z/L)$,  
nondimensionalized by the total stress $FL$.  
The sum $(S+T)/(FL)$ nearly
equals unity for each couple of datapoints, 
confirming that the viscous (solvent) 
stress $\nu_0 U/L^2$ is negligible at the present Reynolds numbers. 
$(b)$ The Reynolds stress $S$ and $(c)$ the polymer stress $T$ 
nondimensionalized by the squared critical velocity $U_{c}^2$:
The full lines are: $S/U_{c}^2=\gamma$ (left, horizontal), 
$S/U_{c}^2=\beta(Re/Re_{c})^2$ (left, oblique), 
$T/U_{c}^2 = 
\delta(Re/Re_{c})^2$ (right). 
The numerical parameters are $\beta=0.15$, 
$\gamma=0.2$ and $\delta=0.02$.}
\label{fig:4} 
\end{figure}

Upon time averaging, eq.~(\ref{eq:1})
reduces to $F_x = - \nu_0 \partial_z^2 \langle{u_x}\rangle + 
\partial_z (\langle{u_x u_z}\rangle - 2 \nu_0 \eta\, 
\langle{\sigma_{xz}}\rangle/\tau)$. 
Utilizing the numerical observation that the Reynolds stress 
${\cal S} = \langle{u_x u_z}\rangle = S \sin(z/L)$ and the polymer stress 
${\cal T} =2 \nu_0 \eta \langle{\sigma_{xz}}\rangle/\tau = - T \sin(z/L)$,
we obtain the momentum budget $F = \nu_0 U/L^2 + S/L + T/L$.
The contribution $\nu_0 U/L^2$ is relevant only in the laminar regime,
and can therefore be neglected. The dependence of the stresses on the rescaled 
Reynolds number is presented in Fig.~\ref{fig:4}. 
Below the threshold the polymer stress is vanishingly small
whereas the Reynolds stress is $S \simeq \beta U^2 \simeq FL$ in agreement with
the observation of a $Re$-independent drag coefficient.
Above $Re_{c}$, the polymer stress makes a significant contribution 
to the momentum budget. 
At the largest $Re$ we can attain, the elastic stress reaches 
almost 50\% of the total stress,
not far from experimental results \cite{PNVH01}.
Rescaling the stresses with the critical velocity squared shows that
above the onset $S$ tends to a constant value $\gamma
U_{c}^2$ (see Fig.~\ref{fig:4}(b)), and the polymer stress
follows the law $T = \delta U^2$ (Fig.~\ref{fig:4}(c)).  
The physical interpretation of these observations is that above the onset 
of drag reduction an increasing fraction of the momentum
injected by the external force is sequestered by polymers,
which are however less effective in absorbing it 
than transverse velocity fluctuations ($\delta < \beta$). This results
in an enhancement of the mean flow with respect to the Newtonian
case, i.e. drag-reduction.  
Inserting the empirical expressions for $S$ and $T$, 
the momentum budget above the onset  reads
$F = \gamma U_{c}^2 / L + \delta U^2/L$, and the 
resulting drag coefficient is
\begin{equation}
f = \left\{
\begin{array}{lcc}
\beta  &
\mbox{for} & Re \lesssim Re_{c}\,,\\
\displaystyle{\gamma \left(\frac{Re_{c}}{Re}\right)^2 + \delta} &
\mbox{for} & Re \gtrsim Re_{c}\;.
\end{array}
\right.
\label{eq:5}
\end{equation} 
This expression is compared with numerical results in Fig.~\ref{fig:3},
where the values of the parameters $\gamma$ and
$\delta$ have been obtained from the data shown in Fig.~\ref{fig:4}.
The agreement is excellent, except possibly for
$Re \approx Re_{c}$, where eq.~(\ref{eq:5}) predicts an abrupt 
transition: from Fig.~\ref{fig:4} this rather appears to be a 
smooth crossover, whose actual shape cannot be 
extracted by means of simple arguments.
The actual values of $\beta$, $\gamma$ and $\delta$ are 
not of utmost importance since they are likely to 
depend on the details of
the driving force, and therefore on the shape of the
velocity profile. What is crucial to drag reduction is that $\delta <
\beta$, or -- in plain words -- that momentum is transferred 
with greater ease to velocity fluctuations than to elastic ones.
Understanding the reasons for this difference would disclose 
the basic physical mechanisms of drag reduction. 

Remarkably, eq.~(\ref{eq:5}) predicts
a maximum drag reduction asymptote (see, e.g., Refs.~\cite{V75,SW00}).
Indeed, by increasing the concentration, drag cannot be reduced below 
the asymptote $f=\delta$, independently of polymer 
elasticity and diffusivity.
In this ultimate regime 
momentum transfer would take place only through polymer stresses. 
However, the present data do not cover a sufficient span of
values of $Re$ to allow us to confirm or reject this prediction.
Numerical simulations at higher resolution should allow to
settle this issue.

We thank M.~Chertkov, B.~Eckhardt, V.~Steinberg, 
and M.~Vergassola for valuable discussions. 
We acknowledge the support of EU under the contract
HPRN-CT-2002-00300 and of MIUR-Cofin 2001023848.
Numerical simulations have been performed at CINECA (INFM parallel
computing initiative).


\begin{thebibliography}{99}

\bibitem{R1883} 
O.~Reynolds, Phil. Trans. R. Soc. {\bf 174}, 935 (1883).

\bibitem{L69} J.~Lumley, Annu. Rev. Fluid Mech. {\bf 1}, 367 (1969). 

\bibitem{V75} P.~S.~Virk, AIChE Journal {\bf 21}, 625 (1975).

\bibitem{MC92} W.~D.~D. Mc Comb, {\it The physics of fluid
    turbulence},
Oxford University Press (1992).

\bibitem{NH95}
R.~H.~Nadolink and W.~W.~Haigh, ASME Appl. Mech. Rev. {\bf 48}, 351 (1995).

\bibitem{SW00}
K.~R.~Sreenivasan and C.~M.~White, J. Fluid Mech. {\bf 409}, 149 (2000).

\bibitem{SBH97}
R.~Sureshkumar {\em et al.\/}, 
Phys. Fluids {\bf 9}, 743 
(1997).

\bibitem{PBNHVH03}
P.~K.~Ptasinski {\em et al.\/}, J. Fluid Mech. {\bf 490}, 251 (2003).

\bibitem{DCLPP03}
E. De Angelis {\em et al.\/}, 
Phys. Rev. E {\bf 67}, 056312 (2003).


\bibitem{DCBP02}
E. De Angelis {\em et al.\/},
http://arxiv.org/nlin.CD/0208016

\bibitem{BCAH87} R.~B.~Bird {\em et al.\/}, 
{\it Dynamics of polymeric fluids} Vol.2, Wiley, New York (1987).



\bibitem{SB95} R.~Sureshkumar and A.~N.~Beris, J. Non-Newtonian
Fluid Mech. {\bf 60}, 53 (1995).

\bibitem{LT88}
T.~S.~Luchik, and W.~G.~Tiederman, J. Fluid Mech. {\bf 190}, 241 (1988).  

\bibitem{WWL87}
W.~W.~Willmarth {\em et al.\/}, Phys. Fluids {\bf 30}, 933 (1987).

\bibitem{profiles}
Sinusoidal mean profiles were first observed in Newtonian ($\eta=0$) turbulence
by V.~Borue, and S.~A.~Orszag,
J. Fluid Mech. {\bf 306}, 293 (1996). See also the related work by 
J.~V.~Shebalin, and S.~L.~Woodruff, 
Phys. Fluids {\bf 9}, 164 (1997).

\bibitem{Rey-num}
It is natural to use the Reynolds number based on solution viscosity:
indeed, for the non-shear-thinning fluid described by 
eqs.~(\ref{eq:1}) and (\ref{eq:2}), 
$\nu$ coincides with the (kinematic) wall viscosity (see, e.g.,
Refs.~\protect{\cite{SBH97,PBNHVH03}}).   
It is also possible to define the equivalent of the 
friction Reynolds number $Re_{\tau}$, often used in experiments:
here, $Re_{\tau}=\sqrt{FL^3/\nu^2}$.
In the Newtonian case $Re_{\tau}\propto Re$.

\bibitem{BFL01} E.~Balkovsky {\em et al.\/}, 
Phys. Rev. E {\bf 64}, 056301 (2001).

\bibitem{timecr} This argument is just a revised
version of the Lumley's ``time criterion'' (see Ref.~\protect\cite{L69}).

\bibitem{LPPT03}
V.~S.~L'vov {\em et al.\/}, 
http://arxiv.org/nlin.CD/0307034.

\bibitem{PNVH01}
P.~K.~Ptasinski {\em et al.\/}, Flow, Turbulence and Combustion {\bf 66}, 159 (2001).



\end{thebibliography}
\end{document}